\documentclass{aa}  

\usepackage{graphicx}
\usepackage{txfonts}
\begin{document}

   \title{Nucleus of active asteroid 358P/Pan-STARRS (P/2012 T1)}

   \titlerunning{Nucleus of active asteroid 358P}

   \subtitle{}

   \author{J. Agarwal\inst{1}
          \and
          M. Mommert\inst{2,3}
          }
   \institute{Max Planck Institute for Solar System Research, Justus-von-Liebig-Weg 3, 37077 G\"ottingen, Germany\\
              \email{agarwal@mps.mpg.de}
   \and
   Lowell Observatory, 1400 W Mars Hill Rd, Flagstaff, AZ 86001, USA
   \and
   Northern Arizona University, Department of Physics and Astronomy, P.O.\ Box 6010, Flagstaff, AZ 86011, USA
             }
   \date{}

 
  \abstract
   {The dust emission from active asteroids is likely driven by collisions, fast rotation, sublimation of embedded ice, and combinations of these. Characterising these processes leads to a better understanding of their respective influence on the evolution of the asteroid population.}
   {We study the role of fast rotation in the active asteroid 358P (P 2012/T1).}
   {We obtained two nights of deep imaging of 358P with SOAR/Goodman and VLT/FORS2. We derived the rotational light curve from time-resolved photometry and searched for large fragments and debris $>$8\,mm in a stacked, ultra-deep image.}
   {The nucleus has an absolute magnitude of $m_R$=19.68, corresponding to a diameter of 530\,m for standard assumptions on the albedo and phase function of a C-type asteroid. We do not detect fragments or debris that would require fast rotation to reduce surface gravity to facilitate their escape. The 10-hour light curve does not show an unambiguous periodicity.}
   {}


   \maketitle
%

\section{Introduction}

Active asteroids have orbits typical of asteroids \citep[with a Tisserand parameter relative to Jupiter ${>}$3, ][]{kresak1972}, but temporarily display dust activity similar to comets \citep{hsieh-jewitt2006,jewitt2012,jewitt_AIV}. While cometary activity is generally thought to be driven by the sublimation of embedded volatiles \citep{whipple1950}, the situation is likely more diverse in asteroids. Main belt asteroids have spent most of the time since their formation between the orbits of Mars and Jupiter, where solar heating prevents the persistence of near-surface ice over solar-system age timescales. But water ice can exist in the interiors of main belt asteroids if protected by a mantle of dust and debris \citep{schorghofer2008,prialnik-rosenberg2009,capria-marchi2012}. 

Currently, five main belt asteroids are known to have ejected dust for extended periods of weeks to months during consecutive perihelion passages \citep{hsieh-jewitt2004,hsieh-meech2011,hsieh_lasagra-3,jewitt_S4-1,hsieh_313P-1,agarwal-jewitt2017}. This combination of protracted and recurrent activity is interpreted as a strong indication that the dust was lifted by gas-drag from sublimating ice \citep{hsieh-jewitt2004}. This subgroup of active asteroids is therefore also called the main belt comets \citep[MBCs;][]{hsieh-jewitt2006}. In some of the other active asteroids, the dust ejection has been an instantaneous process, such as an impact \citep{ishiguro-hanayama2011a,ishiguro-hanayama2011b,kim-ishiguro2017a} or possibly break-up by fast rotation \citep{hirabayashi-scheeres2014a,hirabayashi-scheeres2014b,jewitt_P5-1,drahus-waniak2015}. These same processes are also considered as the most likely triggers of sublimation-driven activity in MBCs by excavating ice from the interior \citep[e.g.][]{haghighipour-maindl2016,haghighipour-maindl2018}. Why sublimation is triggered in some objects and not in others may depend on the abundance and depth of ice, the magnitude of the excavating process, or both. Conversely, fast rotation can support the lifting of dust against gravity by a weak gas flow \citep{jewitt_133P,agarwal-jewitt2016}.

The active asteroid 358P (formerly designated P/2012 T1) was discovered in October 2012 by the Pan-STARRS1 survey \citep{chambers-magnier2016} owing to its bright cometary appearance \citep{wainscoat-hsieh2012}. 358P had passed perihelion at 2.41\,AU on 10 September 2012, about one month before its discovery. The brightness of its dust coma first increased and then decreased over the following months, consistent with sublimation-driven activity \citep{hsieh-kaluna2013}. Numerical simulations reconstructing the production rate, size distribution, and ejection velocity of dust from the appearance of the dust tail indicate that the activity must have been ongoing for 3-5 months, starting at or one month before perihelion \citep{moreno-cabrera2013}. Spectroscopic searches for water vapour yielded an upper limit of 7.63 $\times$ 10$^{25}$ molecules s$^{-1}$ \citep{orourke-snodgrass2013,snodgrass-yang2017}, which is still consistent with weak sublimation of water ice lifting the observed amount of dust. 

We have observed 358P in July and August 2017, at true anomaly angles of 291\degr\ and 296\degr, prior to its return to perihelion in April 2018. The purpose of our observations was to (1) characterise the size, rotation and shape of the bare nucleus, which had been hidden in dust during all previous observations, and (2) to search for a debris trail and larger fragments along the orbit of 358P. The goal was to investigate whether fast rotation might be supporting ice sublimation in lifting the dust from 358P. We describe the observations and data analysis in Section~\ref{sec:obs}. The results are presented in Section~{\ref{sec:results} and discussed in Section~\ref{sec:discussion}.

\section{Observations and data processing}
\label{sec:obs}
\subsection{SOAR/Goodman}

We observed 358P using the Goodman imaging spectrograph
\citep{Clemens2004} mounted on the Southern Astrophysical Research
(SOAR) Telescope located on Cerro Pach\'{o}n, Chile. Observations took
place between 27 July 2017 UT 22:11 and 28 July 2017 UT 10:08; the
observing geometry is described in Table \ref{tab:obs_log}.  We used
Goodman in imaging mode with 2x2 binning (0.3\arcsec pixel scale) and
tracked the telescope at the rate of the target. We used integration
times of 120~s (first 20 frames) and 150~s; background stars were trailed by $l$=3\arcsec\ and $l$=4\arcsec, respectively. The goal of the
observations was the recovery of the target (3$\sigma$ positional
uncertainty of the orbit at the time of observations:
${\sim}3$\arcmin), the measurement of its rotational period using the
broadband VR filter (centred at 610 nm with full width at half maximum (FWHM) of 200 nm), and the measurement of its optical colours using
Sloan g, r, and i filters to put constraints on its taxonomic
classification. Our observations were compromised by highly variable
transparency conditions and cloud coverage, allowing for useful
observations over only about 2~hr out of 9~hr. Furthermore, the target
turned out to be significantly fainter than the brightness predicted
by the JPL Horizons system \citep{giorgini-yeomans1996}. We were able
to recover the target \citep{CBET4425} and extract a partial
light curve from our VR observations. The g, r, and i filter
observations were excluded from further analysis because of their marginal
S/N.

The image data were corrected using bias and dome flat-field images
taken on the same night. Astrometry and instrumental aperture
photometry were derived from the VR images taken during the clear
portion of the night using the PHOTOMETRYPIPELINE
\citep[PP;][]{mommert2017}. The PP provides automated astrometric and
photometric analysis for imaging data and has been specifically
designed for moving target observations. We photometrically calibrated
our observations against Cousins R band transformed from Pan-STARRS
DR1 $r$ magnitudes \citep{tonry-stubbs2012, flewelling-magnier2016}, using no less than
five background stars with solar-like colours
($0.24 \leq g-r \leq 0.64$ and $-0.09 \leq r-i \leq 0.31$ in Sloan
gri) in order to minimise systematic effects caused by the use of the
VR filter while calibrating against R. Because of the highly variable
seeing conditions, we adopted variable apertures for both 358P and the background stars;  the full width at half maximum (FWHM$_{\rm
bg}$) of the background stars varies between 0.5\arcsec and 1.5 \arcsec over the transparent part
of the night. We used aperture radii of $r$=1.3$\times$FWHM$_{\rm bg}$ for 358P, and of $r$=$l$/2 + 1.3$\times$FWHM$_{\rm bg}$ for stars.
In order to mitigate effects caused by the
different aperture sizes, we applied empirical magnitude offsets to
the target photometry in such a way as to keep the flux level of five
bright stars constant; these corrections are significantly smaller
than the typical photometric uncertainties of the target.
The measured photometry from our SOAR observations, which is subject to significant noise, is plotted in the top 
panel of Fig. \ref{fig:lightcurve}. The weighted average brightness of the target from this partial light curve is $m_R= 23.84\pm0.11$. 

\subsection{VLT/FORS2}
\label{sec:vltfors2}
We also observed 358P for 10\,hr beginning on 17 August 2018 UT 23:37, using the FOcal Reducer and low dispersion Spectrograph 2 (FORS2) \citep{appenzeller-fricke1998} mounted on the Unit 1 telescope (UT1) of the European Southern Observatory's (ESO) Very Large Telescope (VLT) on Cerro Paranal in Chile. We obtained 77 images with an exposure time of 400\,s in the R\_SPECIAL+76 filter (central wavelength 655\,nm, bandwidth 165.0\,nm) in 2x2 binning mode, corresponding to a linear pixel scale of 0.25\arcsec. The sky conditions were clear, possibly photometric. The geometry and distance at the time of observation are listed in Table~\ref{tab:obs_log}.

The images were bias-subtracted and flatfielded using the Esorex pipeline, version 3.11.1 \citep{freudling-romaniello2013}. We derived the temporal variation of the seeing
from profiles of 22 field stars not saturating the CCD detector. Stars were trailed by $l$=13.67 pixels because the telescope was tracking the non-siderial motion of 358P, and the trails were inclined by 29$^\circ$ relative to the image $x$-axis. We fitted brightness profiles measured parallel to the $y$-axis with a Gaussian function, and used its FWHM, $D$, as a proxy of that of the seeing disc. The quantity $D$ varied between 3 and 6 pixels (0.75\arcsec\ and 1.5\arcsec) over the course of the night.

\begin{table}
\caption{Journal of our observations of 358P. The quantities $r_h$ and $\Delta$ refer to the heliocentric and geocentric distances in AU, $\alpha$ is the phase angle, and $\nu$ the true anomaly.}             
\label{tab:obs_log}      
\centering                          
\begin{tabular}{l l r r r r}        
\hline\hline                 
Tel. & Start date [UT] & $R_h$ & $\Delta$ & $\alpha$ & $\nu$ \\    
\hline                        
SOAR & 27-07-2017 22:11 & 2.741 & 1.773 & 8.0\degr & 290.8\degr\\
VLT  & 17-08-2017 23:37 & 2.695 & 1.707 & 5.9\degr & 295.7\degr\\
\hline                                   
\end{tabular}
\end{table}

The data analysis procedure used for the VLT data is very similar
to that used for our SOAR observations. Using PP, we performed aperture
photometry on background stars using an aperture radius of $r$=$l$/2 + 1.3$D$ and derived the magnitude zeropoint for
each frame calibrated against Pan-STARRS DR1 $r$ photometry
transformed to Cousins R, requiring a minimum of three non-saturated
background stars with solar-like colours. Target aperture photometry was
obtained using a variable aperture with radius $r$=1.3$D$
(Fig.~\ref{fig:lightcurve}, bottom). We applied empirical magnitude
offsets to the target photometry in such a way as to keep the flux
level of ten bright stars constant; these corrections are
significantly smaller than the typical photometric uncertainties of
the target.
To account for the systematic loss of flux due to the small aperture we corrected the measured magnitudes by subtracting 0.09 (Section \ref{sec:radial_profile}).
The weighted average brightness of the target derived from all VLT data points shown in Fig.~\ref{fig:lightcurve} is $m_R= 23.46\pm0.01$.
\begin{figure}
   \centering
   \includegraphics[width=\columnwidth]{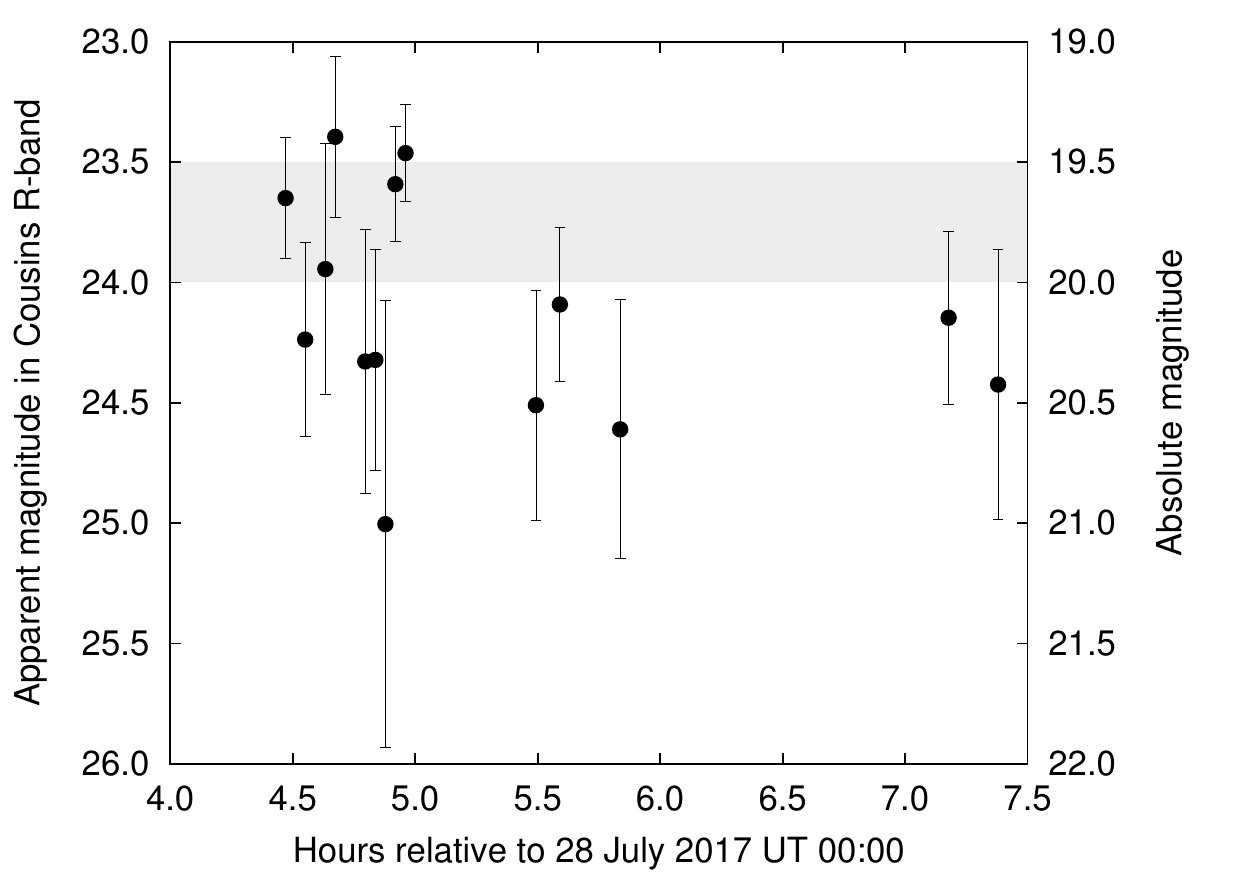}
   \includegraphics[width=\columnwidth]{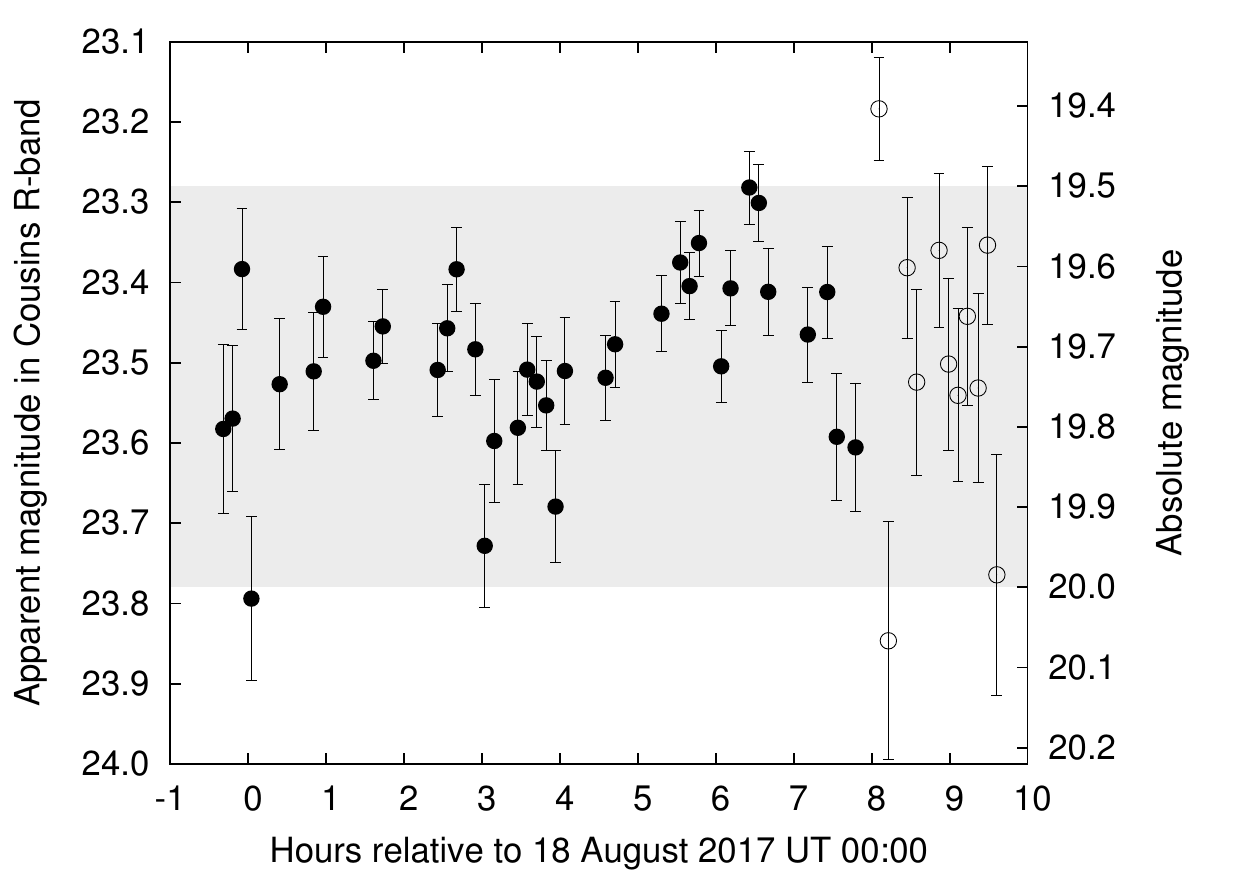}
   \caption{Light curve of 358P on 28 July 2017 observed with SOAR/Goodman (top panel) and on 18 August 2017 observed with VLT/FORS2 (bottom panel). The flux was measured in apertures of radius 1.3$D$, and for the VLT data set corrected by an offset of -0.09 mag to account for the loss of flux due to the small aperture. This correction was not applied for the SOAR data set because the FWHM was measured as average over the moderately trailed stars. The loss of flux due to the small aperture was therefore less than in the FORS data. The correction would have been small compared to the large scatter and error bars of the data. Open symbols in the bottom panel indicate data obtained after UT 8:00 for comparison with Fig.~\ref{fig:mag_vs_seeing}.
The scale on the right-hand side vertical axes indicates the derived absolute magnitude according to Eq.~\ref{eq:absmag} and for $G$=0.15. The shaded areas in both panels indicate the range of absolute magnitudes covered by the VLT data.}
   \label{fig:lightcurve}
 \end{figure}

Figure~\ref{fig:mag_vs_seeing} shows the measured apparent magnitude as a function of the measured FWHM of the seeing, which is proportional to the employed aperture radius. The Pearson correlation coefficient between the corresponding flux and the FWHM is $R$=0.17, indicating a weak correlation between the measured flux and employed aperture size. 
\begin{figure}
   \centering
   \includegraphics[width=\columnwidth]{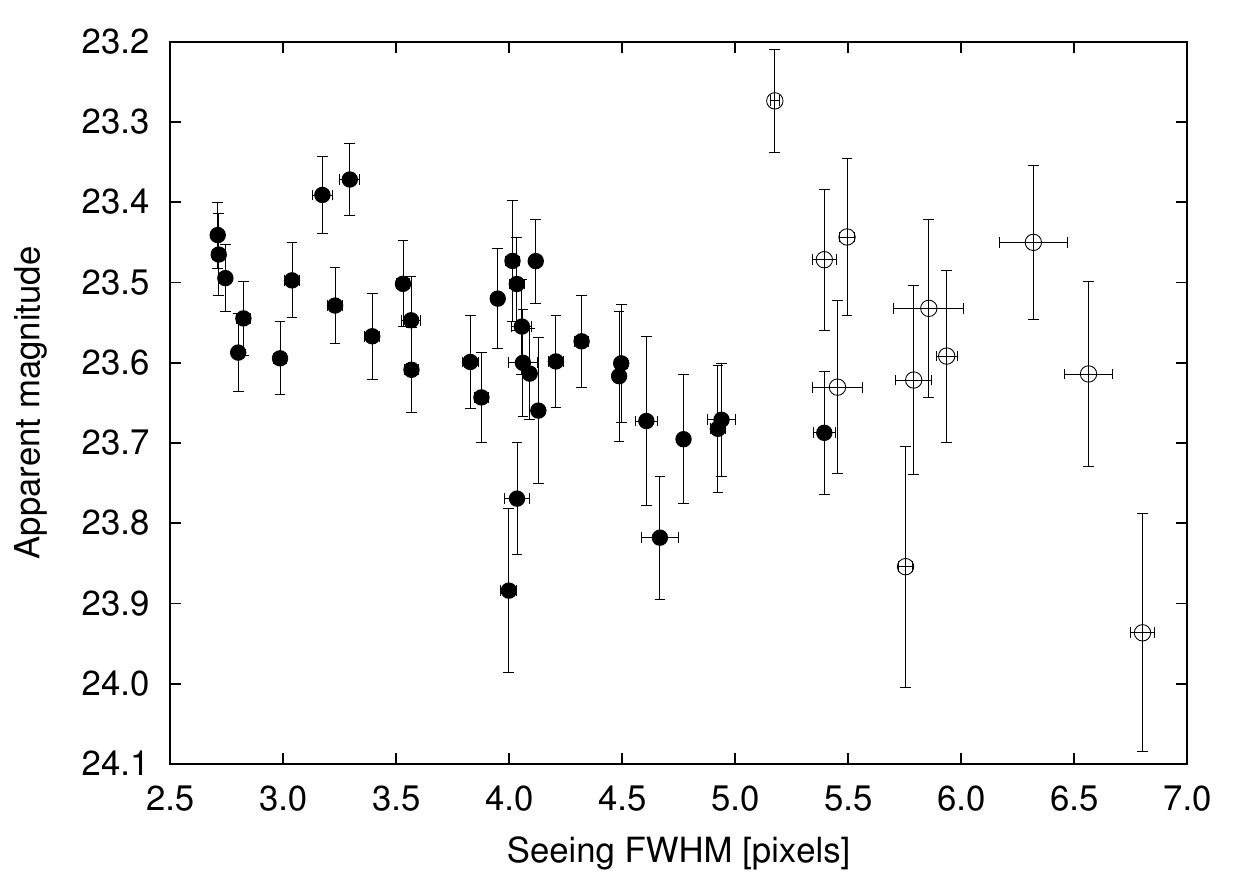}
   \caption{Measured magnitude from VLT data as a function of the seeing FWHM, which is proportional to the aperture radius. The two quantities show a weak degree of statistical dependence ($R$=0.17). Open symbols indicate data points obtained after UT 8:00 (see Fig.~\ref{fig:lightcurve}.)}
   \label{fig:mag_vs_seeing}
 \end{figure}

\section{Results}
\label{sec:results}

\subsection{Light curve}
\label{sec:lightcurve}
To assess the rotational properties of 358P, we concentrate on the VLT
data, which have a longer time coverage than the SOAR data and smaller
photometric uncertainties. Using an implementation of the Lomb-Scargle
periodogram provided by astropy \citep{Astropy2013}, we investigate
periodicity in the brightness variations plotted in
Fig. \ref{fig:lightcurve} (bottom). Since the amplitude of the
variation is of the order of a few 0.1~mag, which is only
slightly higher than the typical photometric uncertainty, we take a
statistical approach. We account for the photometric uncertainties, by
varying the measured brightness in each frame in a Gaussian way,
according to the derived uncertainties. We apply the Lomb-Scargle
algorithm to this randomised data set and derive the power-frequency
distribution for potential light curve periods between 1~hr and
10~hr (the duration of the observations). By repeating this method for
1,000 randomised data sets and summing up the individual
power-frequency distributions, we obtain a distribution that accounts
for the significant photometric uncertainties in the data. Because of the
unknown complexity of the rotational light curve (Fourier order $n$),
we repeat this analysis for a range of Fourier orders
($1 \leq n \leq 5$).

The resulting spectral power distributions are shown in Fig.
\ref{fig:power_frequency}. Relative power maxima independent of the Fourier order $n$ appear at approximately 2~hr and 4~hr. For $n>$1, additional power maxima appear in between, all of which have comparable strength. The 4~hr periodicity, which would indicate an 8~hr rotation period for a double-peaked light curve, agrees with the light curve
behaviour observed during the first 8~hr of our VLT/FORS2 observations (filled symbols in Figs.\ref{fig:lightcurve} and \ref{fig:mag_vs_seeing}). Data from the last two hours (open symbols in Figs.\ref{fig:lightcurve} and \ref{fig:mag_vs_seeing}) were obtained under deteriorated seeing conditions and may be less reliable.
From Fig.~\ref{fig:power_frequency} we are unable to derive an unambiguous rotational period for 358P. Potential explanations
are that the object is rather spherical or observed from a polar perspective, leading to a light curve
amplitude that is smaller than our photometric uncertatinties, or that
the object's rotational period is significantly longer than 10~hr. We
discuss this result in detail in Section \ref{sec:discussion}.

 \begin{figure}
   \centering
   \includegraphics[width=\columnwidth]{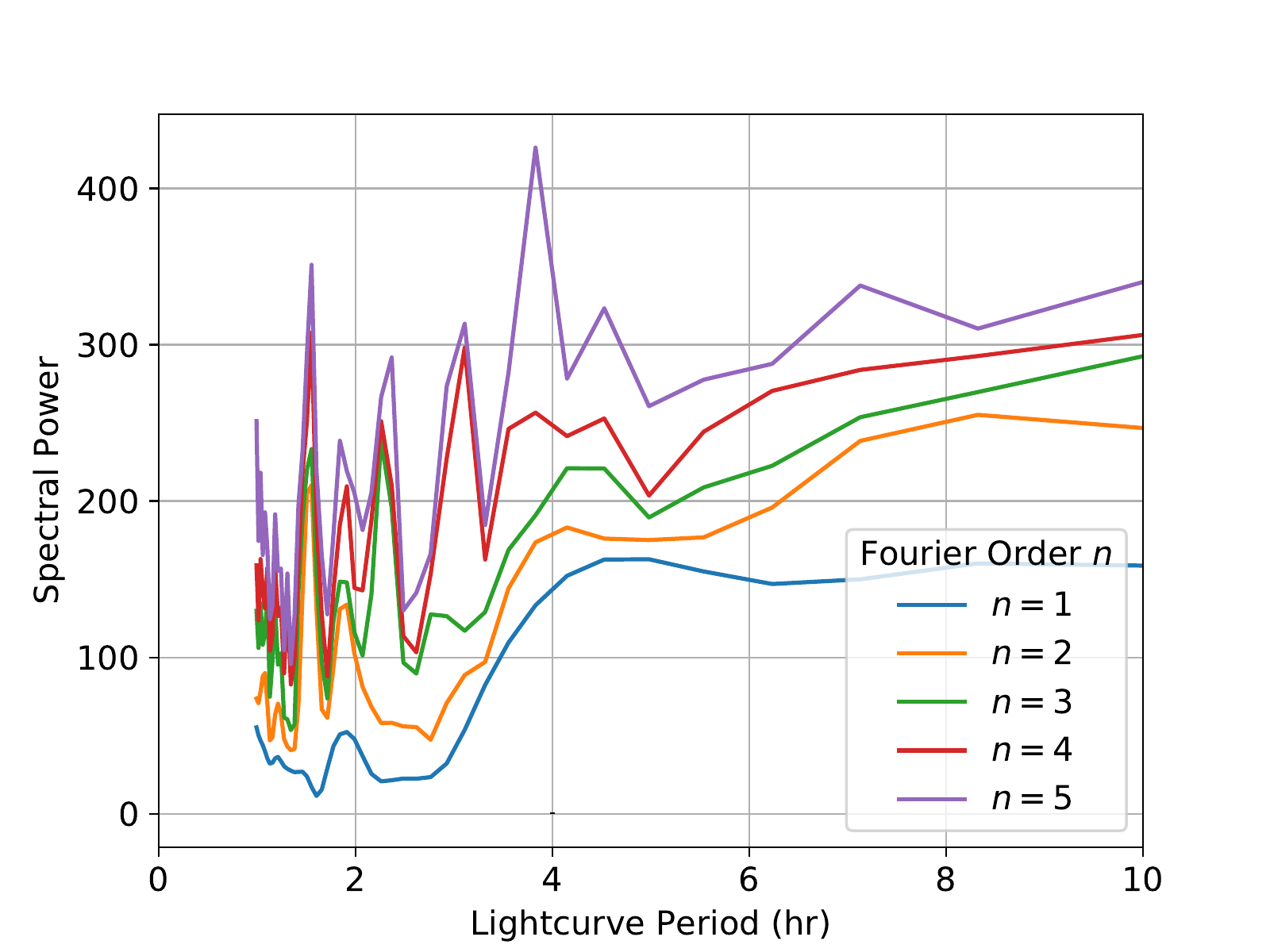}
   \caption{Spectral power distributions for the light curve period (which corresponds to half the rotation period for a shape-induced double peaked light curve) of
     358P using a statistical approach. For any period in the range,
     the distributions provide the spectral power as derived with a
     Lomb-Scargle algorithm; the coloured lines denote different
     Fourier orders. 
     We are unable to identify a clear periodicity in the period range that was
     probed.}
   \label{fig:power_frequency}
 \end{figure}

\subsection{Radial profile and dust coma}
\label{sec:radial_profile}

To search for dust near the nucleus in the FORS2 images, we compared radial flux profiles of 358P with those of field stars. Since stars were trailed in the exposures targeting 358P, we used two calibration exposures of standard star fields obtained at siderial tracking at the beginning and end of the night, respectively, to measure the stellar radial profiles. We averaged the normalised profiles of 14 and 26 non-saturated field stars, respectively. For comparison, we selected exposures of 358P obtained under similar seeing conditions (as indicated by the FITS header keyword FWHMLINOBS). 
The keyword FWHMLINOBS seems to have generic values for the first 21 frames of the data, which we consider unreliable. For the remaining 56 frames, the values of FWHMLINOBS and our measured $D$ are well correlated with a Pearson correlation coefficient of 0.88. The frames we used for comparison to the standard star fields are from this second part of the data set.
We used a single 358P exposure (frame 40) for comparison with the first standard star field, and six exposures (frames 70 + 73-77) for the second star field that we averaged in the co-moving frame. Fig.~\ref{fig:radial_profiles} shows the resulting radial profiles. For both values of the seeing parameter $D$, the profiles of 358P and the stars are similar, and they are also similar for different seeing conditions if expressed in units of $D$. The radial profile of 358P is consistent with that of a point source. It does not show evidence of broadening by dust. Fig.~\ref{fig:radial_profiles} shows that the percentage of flux enclosed in a given multiple of $D$ is independent of the value of $D$. At a radius of 1.3$D$, the aperture encloses 92\% of the total flux, requiring a magnitude correction of $\Delta M$=-0.09 (cf. Fig.~\ref{fig:lightcurve}).
\begin{figure}[h]
   \centering
   \includegraphics[width=\columnwidth]{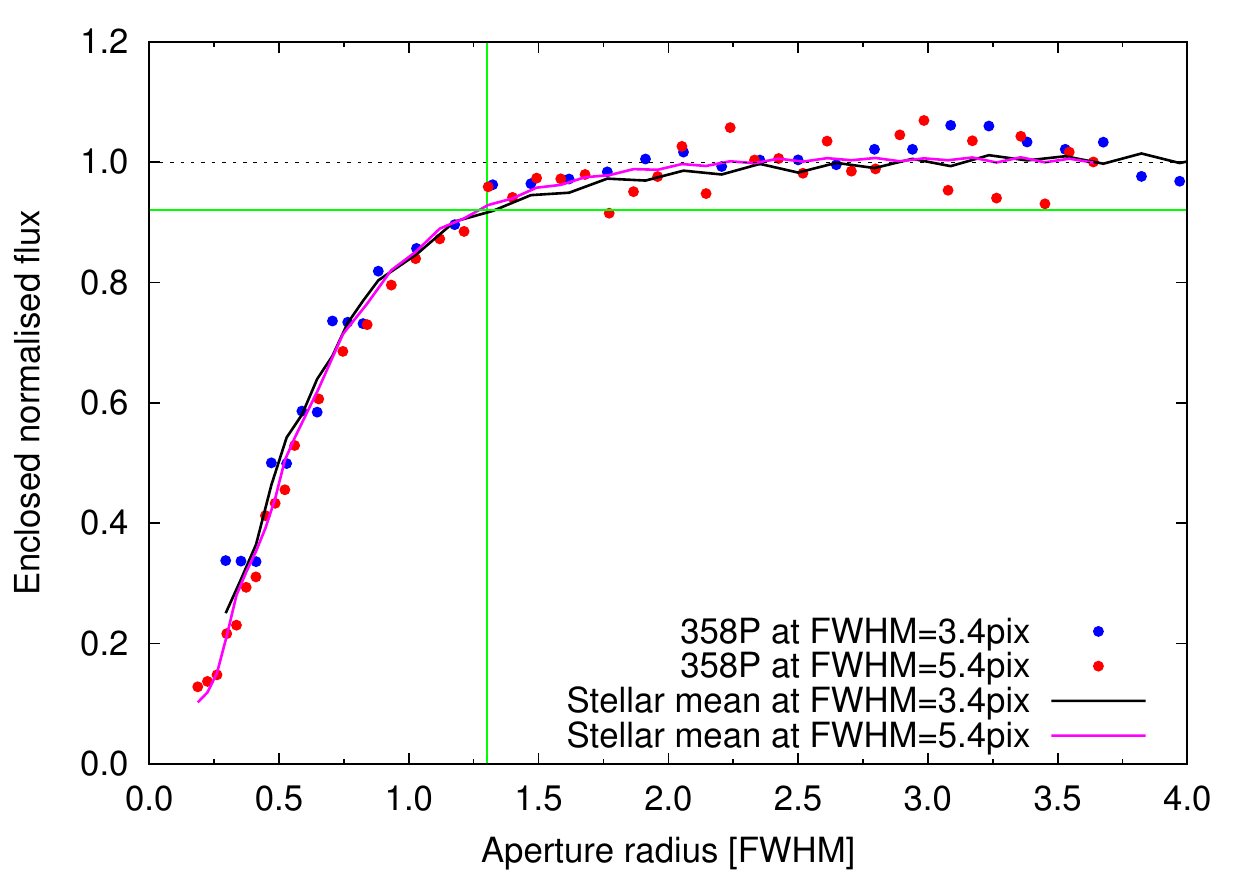}
   \caption{Normalised radial profiles of 358P and field stars in the FORS2 data, measured for two different values of the seeing FWHM, $D$. The green lines indicate the aperture radius in which the light curve (Fig.~\ref{fig:lightcurve}) was measured and the corresponding flux level of 92\%. 
The x-axis is in units of the FWHM at observation, i.e. a unit interval corresponds to 3.4 pixels for the blue dots and 5.4 pixels for the red dots.}
   \label{fig:radial_profiles}
 \end{figure}

\subsection{Nucleus size}

We constrain the size of the nucleus from the
VLT/FORS2 observations only, since they cover a longer fraction of
the target's light curve and we assume these observations provide a better approximation of
its mean brightness than the SOAR/Goodman observations. 
The absolute magnitude, $H_R$, (corresponding to
$r_h$=$\Delta$=1\,AU and $\alpha$=0) is given by
\begin{equation}
H_R = m_R - 5\log_{10}(R\Delta) + 2.5\log_{10}(\Phi(\alpha)),
\label{eq:absmag}
\end{equation}
where $\Phi(\alpha)$ is the phase function describing the ratio of the scattered light at phase angle $\alpha$ to that at $\alpha$=0$^\circ$. We used the HG approximation of $\Phi(\alpha)$ with $G=$0.15 corresponding to C-type asteroids \citep{bowell-hapke1989}. The measured apparent magnitude of $m_R$=23.46$\pm$0.01 (Section \ref{sec:vltfors2}) corresponds to an absolute magnitude of $H_R$=19.68$\pm$0.01. Assuming an S-type phase function with $G$=0.25 would result in $H_R$=19.74$\pm$0.01 instead. The absolute magnitude derived from the average brightness in the SOAR/Goodman data ($m_R$=23.84$\pm$0.11) is $H_R$=19.84$\pm$0.11 ($G$=0.15) or $H_R$=19.91$\pm$0.11 ($G$=0.25). Since the stated uncertainties of the mean correspond to 1$\sigma$, the two measurements are marginally consistent.

We base our following estimate of the nucleus size on the value $H_R$=19.68$\pm$0.01 measured from the VLT data and assuming a C-type phase function. For a geometric albedo of $p_V$=0.06  (we approximate $p_R$ with $p_V$, which is  reasonably close for low albedos)\ typical for C-type asteroids \citep{nugent-mainzer2016}, the absolute magnitude corresponds to an equivalent-sphere radius of 
\begin{equation}
r_n = p_V^{-1/2} 10^{(M_\odot - H_R)/5} \times 1\,{\rm AU}, 
\label{eq:diameter}
\end{equation}
\citep[e.g.][]{harris-lagerros2002}. For a solar magnitude of $M_\odot=-27.15$ in Cousins R-band \citep{binney-merrifield1998}, we obtain $r_n$=263\,m. The main source of uncertainty is the unknown albedo of 358P. The C-type albedos given in \citet{nugent-mainzer2016} vary within a factor of 3, which leads to a radius uncertainty of a factor 1.7. The unknown phase function, by contrast, induces only a small uncertainty due to the low phase angle at the time of observation. The radius derived for $G=0.25$ would only be 3\% smaller than for $G=0.15$. The radius uncertainty introduced by the photometric uncertainty is also 3\%.
The escape speed from the surface of a non-rotating body of sub-km size and assuming a density of $\rho$=1500\,kg\,m$^{-3}$ \citep{hanus-viikinkoski2017} is $v_{esc} = 0.2$\,m\,s$^{-1}$. The density reported by \citet{hanus-viikinkoski2017} was measured for C-type asteroids $>$100 km in diameter. We use this value for the much smaller 358P owing to the lack of measurements for a C-type object of this size. An asteroid with known density and comparable in size to 358P is the S-type (25143) Itokawa. Its density $\rho$=1900\,kg\,m$^{-3}$ \citep{fujiwara-kawaguchi2006} is near the lower end of the interval of (2000 -- 4000) \,kg\,m$^{-3}$ observed for S-types by \citet{hanus-viikinkoski2017}, which is a result of Itokawa's high macroporosity and rubble-pile nature \citep{fujiwara-kawaguchi2006}.

\subsection{Upper limits for trail brightness and fragment size}

We searched for a faint dust trail in a deep composite image of all 77 FORS2 exposures. To obtain the composite, we first scaled each background-subtracted exposure, $i$, with a factor $f_i = 10^{0.4 \Delta M_i}$ to compensate for the variable atmospheric extinction, where $\Delta M_i$ is the offset of the averaged instrumental magnitude of a set of field stars from an (arbitrary) reference magnitude. Subsequently, we averaged all frames in the siderial reference frame rejecting the faintest and the three brightest values at each position, and subtracted the resulting stellar composite from each exposure. Finally, we averaged all frames in the co-moving frame of 358P with the same rejection rule. Figs.~\ref{fig:full_image} and \ref{fig:image} show the resulting deep image of 358P.
\begin{figure*}[t]
   \centering
   \includegraphics[width=\textwidth]{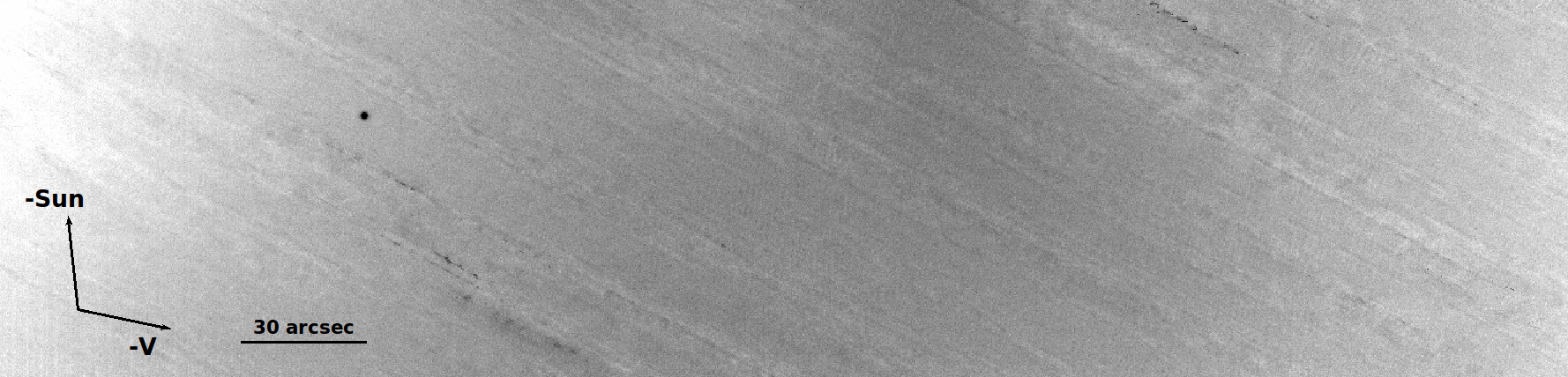}
   \caption{358P in a composite of 77 FORS2 exposures, averaged in the co-moving frame of the asteroid. The brightness scale is inverted and linear, and the arrows indicate the anti-solar direction and the projected negative orbital velocity vector. No indication of a debris trail (which should be parallel to the negative velocity vector) is visible.}
   \label{fig:full_image}
 \end{figure*}
\begin{figure}[h]
   \centering
   \includegraphics[width=\columnwidth]{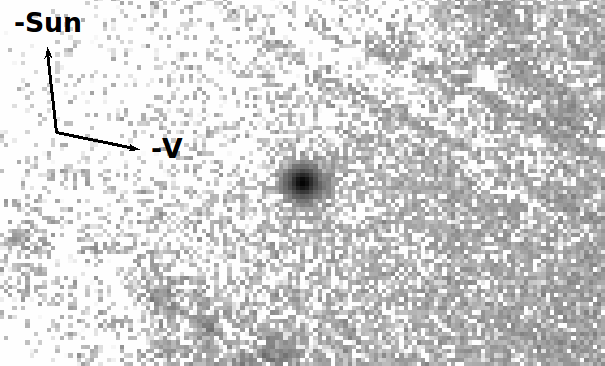}
   \caption{Zoom-in on Fig.~\ref{fig:full_image}. The brightness scale is inverted and logarithmic, and the size of the image is 35\arcsec\ $\times$ 21\arcsec.}
   \label{fig:image}
 \end{figure}
To derive an upper limit for the surface brightness of the debris trail, we assume that an extended linear object having S/N=1 per pixel is easily detectable to the human eye \citep{agarwal-mueller2010}, and that the trail surface brightness must therefore be smaller than the local standard deviation of the background flux, $\sigma$=20\,ADU, in the composite image (Fig.~\ref{fig:full_image}). This corresponds to an upper limit on the surface brightness of 28.4 mag/arcsec$^{2}$ in Cousins R, which is three magnitudes fainter than the surface brightness of the debris trails of the active asteroids P/2010 A2 \citep{jewitt-ishiguro2013} and 331P \citep{drahus-waniak2015}.

We proceed to infer an upper limit on the size of individual large fragments assuming that point sources with $S/N>$3 would be detectable. Following \citet{makovoz-marleau2005}, we use $S/N = f / (\sigma \sqrt{\pi} r)$, where $f$ is the total flux from the point source and $r$ is the radius of the aperture. For $r$ = 4 pixels we find that $f>425$\,ADU is required, corresponding to a limiting apparent magnitude of 27.8 in Cousins R-band. At the position of 358P during our VLT/FORS2 observations, this corresponds to an absolute magnitude of 24.0 (Eq.~\ref{eq:absmag}) or a diameter of 72\,m (Eq.~\ref{eq:diameter}).

\subsection{Production of cm-sized debris}
Whether a particle ejected in 2012 remains in the FOV of our 2017 observation depends on its velocity component, $v_e$, parallel to the orbital motion of 358P upon decoupling from its gravitational influence \citep{mueller-green2001}, and on the ratio of solar radiation pressure to local solar gravity, $\beta = 5.77 \times 10^{-4} Q_{\rm pr}/(\rho a)$, where $a$ and $\rho$ are the radius and bulk density of the particle and the dimensionless parameter $Q_{\rm pr}$ characterises the optical properties of the material \citep{burns-lamy1979}. We numerically simulated the motion of test particles ejected during the 2012 perihelion passage that have a wide range of values for $\beta$ and $v_e$, and calculated their positions relative to 358P at the time of our FORS2 observation in August 2017. Only particles that have
\begin{equation}
\beta < (7.9\,\mathrm{s m^{-1}} v_b + 3.9) \times 10^{-5}  
\end{equation}
would still be in the FOV of our observations, where $v_b = -v_e$ is positive towards the direction opposite to the orbital motion of 358P. Particles ejected to this direction stay closer to the nucleus than particles ejected to the forward direction at the same relative speed because backward ejection decreases the orbital energy and period of the particle, counteracting radiation pressure. The hatched blue area in Fig.~\ref{fig:velocity} shows possible backward ejection speeds as a function of particle size. 

\begin{figure}[h]
   \centering
   \includegraphics[width=\columnwidth]{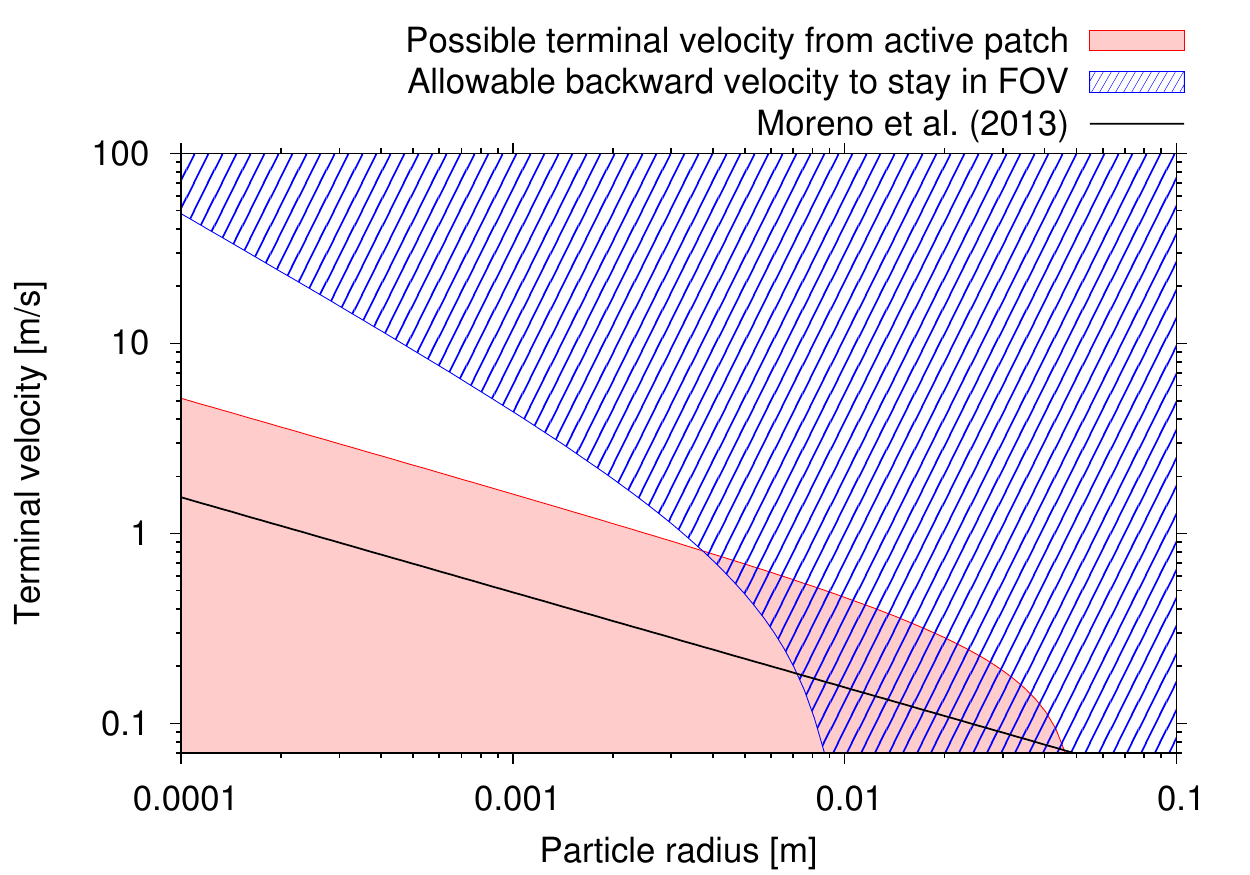}
   \caption{Possible terminal velocities as a function of particle size. The blue hatched area shows velocities towards the negative orbital direction that would enable a particle of given size to remain in the FOV of FORS2 in August 2017. The red area indicates velocities consistent with the observed upper limit of the gas production rate according to Eq.~A5 in \cite{jewitt_133P} and for the parameters described in the text. The overlap region corresponds to combinations of size and backward ejection velocity for particles that could be expected in our image. The solid line represents the velocity-size relationship from \citet{moreno-cabrera2013} given in Eq.~\ref{eq:v_beta} assuming a bulk density of 1500\,kg\,m$^{-3}$ and $v_0$=25\,m\,s$^{-1}$.}
   \label{fig:velocity}
 \end{figure}

In the following, we derive the maximum particle speed consistent with the measured upper limit of the water production rate $Q_{max}$=7.63$\times$10$^{25}$ molecules\,s$^{-1}$ \citep{orourke-snodgrass2013}, assuming that the activity was confined to a patch on the surface. The temperature of a sublimating surface containing water ice at the heliocentric distance $r_h$ is given from the equilibrium of solar irradiation with cooling by radiation and sublimation,
\begin{equation}
\frac{L}{N_A m_{H_2O}} Q_{H_2O} + \epsilon \sigma T^4 = (1 - A_B) \frac{I_\odot}{r_h^2} \cos \theta,
\end{equation}
where $\epsilon$ and $A_B$ are the emissivity and Bond albedo of the surface, $\sigma$ and $N_A$ the Stefan-Boltzmann and Avogadro constants, $L$ = 51000 J/mol the latent heat of water ice, $m_{H_2O}$ the molecular mass of water, and $\theta$ the angle between the surface normal and solar direction. The value $Q_{H_2O}$ is the sublimation rate of water ice in vacuum in kg\,s$^{-1}$\,m$^{-2}$ given by 
\begin{equation}
Q_{H_2O} = p_{subl}(T) \sqrt{\frac{m_{H_2O}}{2 \pi k_B T}},
\label{eq:QM}
\end{equation}
where the sublimation pressure is given by $p_{subl}(T) = A \exp({-B/T})$ with $A$ = 3.56$\times$10$^{12}$\,Pa and $B$ = 6141\,K \citep{fanale-salvail1984}.

Assuming extreme values of $\epsilon$=1, $A_B$=0 and normal incidence ($\theta$=0), we obtain an upper limit for the temperature $T_{max}$=191\,K at the perihelion distance of 358P ($r_h$=2.41\,AU), and a sublimation rate of $Q_{H_2O}/m_{H_2O}$=1.74$\times$10$^{21}$ molecules\,s$^{-1}$\,m$^{-2}$. Assuming that the activity was confined to a circular patch on the surface, and that this patch was not illuminated and inactive for 50\% of each diurnal cycle, the observational upper limit $Q_{max}$ translates to a maximum patch radius of 170\,m, or 10\% of the surface. 

From a patch of this size, assuming a bulk density of 1500\,kg\,m$^{-3}$ for both dust and nucleus and a unit gas drag coefficient, the maximum escaping grain size would be 5\,cm \citep[Eq. A6 in][cf. Fig.~\ref{fig:velocity}]{jewitt_133P}. The maximum liftable grain radius scales about linearly with the patch radius. Assuming instead a comet-like density of 500\,kg\,m$^{-3}$ for both dust and nucleus would increase the maximum liftable grain radius to $\sim$35\,cm.

The size- and velocity ranges compatible with both the gas production and dynamical constraint (overlapping red and hatched areas in Fig.~\ref{fig:velocity}) are consistent with the velocity-size relation used by \citet{moreno-cabrera2013}, that is 
\begin{equation}
v = v_0 \, \beta^{0.5}
\label{eq:v_beta}
.\end{equation}
where $v_0$ = 25\,m\,s$^{-1}$. Fig.~\ref{fig:velocity} indicates that our observations are sensitive to particles in the size range 8 mm$<\!a\!<$5 cm.

Adopting Eq.~\ref{eq:v_beta} and assuming isotropic ejection, we derive an upper limit on the production rate of such particles from our detection limit of the trail surface brightness. We found that the 2017 position of particles depends only weakly on the actual emission time within an interval of a few months around perihelion, and therefore studied only the motion of grains ejected at perihelion. 
We calculated the image position as a function of size and the corresponding surface brightness for a given total dust production using the computer code described in \citet{agarwal-mueller2010}. 

For differential size distribution exponents $\alpha\!>$-4.5, the surface brightness has a maximum near the eastern end of the trail, while for $\alpha\!<$-4.5, it increases towards the west. For $\alpha$=-3.5, the maximum surface brightness would be at distances between 70\arcsec\ and 100\arcsec.
Particles with $a\!>$8\,mm would be located $>$20\arcsec\ west of the nucleus. 

Assuming $\alpha$=-3.5 and the same phase function and albedo as for 358P, our detection limit of 28.4\,mag/arcsec$^2$ corresponds to an upper limit of 52$\times$10$^6$\,kg of particles having 8\,mm$<\!a\!<$5\,cm ejected over the whole 2012 period of activity. The total R-band magnitude of particles in the 2017 FORS2 FOV would have been $>M_{min}$=21.7, and their velocities would have been of the order 20\,cm\,s$^{-1}$ (Fig.~\ref{fig:velocity}).
They would have remained within an aperture of 3\arcsec\ for six months in 2012 (1\arcsec\ $\sim$ 1000\,km). 
We derive an upper limit of $(A\!f\!\rho)_{FOV}$=1\,cm \citep{ahearn-millis1995a} contributed by particles acceptable in the 2017 FOV to the $A\!f\!\rho_{obs}\sim$12\,cm measured within an aperture of 5\arcsec\ \citep{hsieh-kaluna2013}.  

Our observations imply that the contribution of particles with $a\!>$8\,mm to the coma brightness observed in 2012 must have been small, while we cannot constrain the production rates of particles with $a\!<$8\,mm. 
The model described by \citet{moreno-cabrera2013} implies a mass of $\sim$15$\times$10$^6$\,kg of particles in the 1 -- 10\,cm size range, which corresponds to $A\!f\!\rho_m \sim$0.3\,cm; this value is consistent with our upper limit, but represents 75\% of the total produced mass. This shows that despite their low surface brightness, such particles could have carried a significant percentage of the ejected dust mass.

\section{Summary and discussion}
\label{sec:discussion}
We observed the active asteroid 358P at true anomaly angles of 290.8\degr\ and 295.7\degr\ and obtain the following key results:
\begin{itemize}
\item The peak-to-peak amplitude of the rotational light curve in August 2017 was likely $\sim$0.2\,mag, but might be smaller than that.
\item The 10 hr VLT/FORS2 light curve observation does not show an obvious periodicity but might marginally indicate a rotation period of $\sim$8 hrs. 
\item The radial profile of 358P in August 2017 was consistent with that of a point source, showing no evidence of broadening by dust.
\item We derive an average absolute magnitude $H_R$=19.68$\pm$0.01 assuming a C-type phase function, and $H_R$=19.74$\pm$0.01 for an S-type phase function.
\item Assuming a geometric albedo of 0.06, this corresponds to a cross-section-equivalent sphere of 530\,m diameter that has an uncertainty of a factor 1.7 owing to the unknown albedo.
\item For a density of 1500\,kg\,m$^{-3}$, the surface escape velocity is 0.2\,m\,s$^{-1}$.
\item Our observation was sensitive to individual fragments $>$70\,m in diameter, which we did not detect.
\item We did not detect a debris trail along the projected orbit, and derive an upper limit to its surface brightness in Cousins R-band of 28.4\,mag/arcsec$^2$, three magnitudes fainter than for active asteroids with known debris trails.
\item Our observation is sensitive to dust particles in the size range 8 mm -- 5 cm, for which we derive an upper limit of 52$\times$10$^6$\,kg produced during the  2012 perihelion passage.
\item The contribution of such particles to the coma brightness in 2012 must have been $<$10\%, while they may still have carried a significant percentage of the ejected mass. 
\end{itemize}
Our primary goal was to study the rotation state of 358P and its
possible inter-relation with the activity. We approached this
question both directly by measuring the rotational light curve, and
indirectly through the size and velocities of the ejected dust.

The latter approach did not yield any indication that fast rotation
would be needed to explain the ejection of refractory material. We do
not detect large fragments or debris $>$8\,mm, and the derived
upper limits are consistent with acceleration by gas drag only. 

The direct measurement of the rotational light curve was limited by the
unexpected faintness of 358P and the likely small amplitude of the light curve. Our attempt at deriving the rotation period of the target was
inconclusive. While a rotation period of ${\sim}$8~hr seems compatible 
with some of our data, it is incompatible
with other parts obtained at unfavourable seeing. Our ability to derive a conclusive rotational period
might be hampered by the fact that the actual light curve
amplitude of the target is smaller than the variability that we
measured. This possible explanation is underlined by the fact that our
photometric uncertainties are of the same order of magnitude. A
different possible explanation for our failing to derive the
rotational period might be that the rotation is insufficiently
sampled. The period might be much longer than the 10~hr interval for which we
observed 358P, or it might be small compared to our cadence (${\sim}$10~min). The latter seems unlikely because it would imply that 358P rotates faster than the critical 2.2\,hr period applicable for rubble piles, although its size indicates a high probability that 358P is a rubble pile \citep{warner-harris2009}. 

We find some indication of a discrepancy between the average absolute brightness of 358P during our SOAR/Goodman and VLT/FORS2 observations (Section
   \ref{sec:obs}). A significant variability in brightness is also seen in additional photometry obtained with Gemini South \citep{CBET4425}, where the corresponding absolute magnitudes fluctuate of the order of 1~mag over the course of several weeks.
This might support the idea that the light curve
 period is longer than the duration of our observation (10 hours) or that
 358P deviates from a simple periodicity owing to presently
 unknown reasons. Possible factors adding complexity to the light curve
 could be the existence of a binary partner, or non-principal axis
 rotation (``tumbling''). 
We conclude that we cannot rule out a rotation faster than 10~hr if the
light curve amplitude (peak-to-peak) is less than ${\sim}$0.2~mag, nor a slow rotation with a period longer than 10~hr.

\begin{acknowledgements}
\\
JA was supported in part by the European Research Council (ERC) Starting Grant No. 757390.
MM was supported in part by NASA grant No.\ NNX17AG88G from the
Near Earth Object Observations programme.
\\
This work is based in part on observations obtained at the Southern Astrophysical Research (SOAR) telescope, which is a joint project of the Minist\'{e}rio da Ci\^{e}ncia, Tecnologia, Inova\c{c}\~{a}os e Comunica\c{c}\~{a}oes (MCTIC) do Brasil, the U.S. National Optical Astronomy Observatory (NOAO), the University of North Carolina at Chapel Hill (UNC), and Michigan State University (MSU).
\\
The results presented in this work are based in part on observations made with ESO Telescopes at the La Silla Paranal Observatory under programme ID 099.C-0530(A).
\\
We thank the night astronomer Jesus M. Corral-Santana and the VLT staff for their outstanding work in carrying out the FORS2 observations. We thank the referee for the comments that significantly helped to improve the manuscript.
\\
This research has made use of NASA's Astrophysics Data System Bibliographic Services.

\end{acknowledgements}

   \bibliographystyle{../bibtex/aa} 
   \bibliography{/home/agarwal/Latex/refs.bib} 

\end{document}